\documentclass[12pt]{article}

\begin{document}
\begin{center}
{\bf EFFECTIVE ACTION AND ELECTROMAGNETIC POLARIZABILITIES OF NUCLEONS IN QCD STRING THEORY}\\
\vspace{5mm}
 S.I. Kruglov \footnote{E-mail: krouglov@utsc.utoronto.ca}\\
\vspace{5mm}
\textit{University of Toronto at Scarborough,\\ Physical and Environmental Sciences Department, \\
1265 Military Trail, Toronto, Ontario, Canada M1C 1A4}
\end{center}

\begin{abstract}
The effective action for baryons in the external electromagnetic
fields is obtained on the basis of the QCD string theory. The area
law of the Wilson loop integral, the approximation of the
Nambu-Goto straight-line strings, and the asymmetric quark-diquark
structure of nucleons are used to simplify the problem. The
spin-orbit and spin-spin interactions of quarks are treated as a
perturbation. With the help of the virial theorem we estimate the
mean radii of nucleons in terms of the string tension and the Airy
function zeros. On the basis of the perturbation theory in small
external electromagnetic fields we derive the electromagnetic
polarizabilities of nucleons. The electric and diamagnetic
polarizabilities of a proton are $\bar{\alpha }_p= 10\times
10^{-4}~ fm^3$, $\beta _p^{dia}=-8\times 10^{-4}~ fm^3$ and for a
neutron we find $\bar{\alpha }_n=4.2\times 10^{-4}~ fm^3 $, $\beta
_n^{dia}= -5.4\times 10^{-4}~ fm^3$. Reasonable values of the
magnetic polarizabilities $\bar{\beta}_p=(5\pm 3)\times 10^{-4}~
fm^3$, $\bar{\beta}_n=(7.6\pm 3)\times 10^{-4}~ fm^3$ are
estimated using the $\Delta$ contribution to the paramagnetic
polarizability of the nucleons.
\end{abstract}

\section {Introduction}

In recent papers [1,2] we have calculated some electromagnetic
characteristics of mesons in the framework of the QCD string theory [3,4].
Starting with QCD and excepting the area low for the Wilson loop integral,
the effective action for mesons in external electromagnetic fields was
derived. This approach takes into account the chiral symmetry breaking and
the confinement of quarks [3,4]. This work is the continuation of [5] on the
case of baryons, i.e. the three-quark system. We make some assumptions such
as (i) the area law behavior of the Wilson loop, (ii) the straight-line
approximation of the strings, (iii) the handling of spin degrees of freedom
as a perturbation, (iiii) the quark-diquark structure of baryons. The
spin-orbit, spin-spin interactions and Coulomb like short-range
contributions are also neglected, as a first approximation. The area-law of
the Wilson loop integral (or the confinement of quarks) was confirmed by the
Monte-Carlo simulations and by the theoretical considerations. One of the
approaches which explains the confinement of quarks is the approach [6]
introducing stochastic gluon vacuum fields. The second one uses the dual
Meissner effect to guarantee the confinement of colour [7,8]. In our
consideration, the confinement of quarks is postulated by excepting the
area-law of the Wilson loop. The use of the straight-line approximation for
the Nambu-Goto string is useful to simplify the calculations. Neglecting
short-range interactions is justified by considering large distances. The
quark-diquark structure of baryons was considered in the framework of the
nonrelativistic [9] and relativistic [10] models. In this case the linear
baryon Regge trajectories have the same slope as for mesons [3].

The paper is organized as follows. In Section 2 we describe the general
background and derive the Green function of a baryon. The effective action for
baryons in external electromagnetic fields based on the proper time method
and Feynman path-integrals is found in Section 3. Section 4 contains the
calculation of average distances between quarks and electric
polarizabilities of nucleons. We derive diamagnetic polarizabilities of
protons and neutrons using the perturbative expansion in the small magnetic
fields in Section 5. Section 6 contains a conclusion.

\section {The Green function of three-quark system}

Here we derive the Green function of three-quark system using the Schwinger
proper time method and the Feynman path-integrals. Our goal is to calculate
some electromagnetic characteristics of nucleons. For this purpose we need
the effective action for baryons in external electromagnetic fields. The
method of the Green functions will be explored. Let us consider the Lorentz-covariant
and gauge invariant combination of three-quark, colourless system (baryon) [4]
\begin{equation}
X_B(x,y,z,C_i)=\epsilon _{abc}\left[ \Phi (Z_0,x)q(x)\right] _a\left[ \Phi
(Z_0,y)q(y)\right] _b\left[ \Phi (Z_0,z)q(z)\right] _c,  \label{(1)}
\end{equation}
where $q(x)$ is a quark bispinor; $a$, $b$, $c$ are colour indexes so that $
\left[ \Phi (Z_0,x)q(x)\right] _a=\Phi _{aa^{^{\prime
}}}(Z_0,x)q_{a^{^{\prime }}}(x)$ and $\epsilon _{abc}$ is the Levi-Civita
symbol ($\epsilon _{123}=1$). We imply that quark fields $q(x)$, $q(y)$ and $
q(z)$ possess the definite flavours which will be specified later at the
considering nucleons. As usual there is a summation on repeating indexes.
The gauge invariance is guaranteed here by introducing the parallel
transporter [4]:
\begin{equation}
\Phi (Z_0,x)=P\exp \left\{ ig\int_x^{Z_0}A_\mu dz_\mu \right\} ,
\label{(2)}
\end{equation}
where $P$ is the ordering operator along the contour $C_1$ of integration, $
g $ is the coupling constant, $A_\mu =A_\mu ^a\lambda ^a$; $A_\mu ^a$ are
the gluonic fields; $\lambda ^a$ are the Gell-Mann matrices and $Z_0$ is the
string junction. Although we have an attractive feature - gauge invariance, the
function (1) depends on the form of the contour $C$.
As $X_B(x,y,z,C_i)$ is a gauge invariant object, it obeys
the Gauss law on the spacelike surface $\Sigma $. Three quarks (with fields $
q(x)$, $q(y)$ and $q(z)$) are situated in the four-points $x$, $y$, $z$,
respectively and connected with the four-point $Z_0$ which is arbitrary
four-point. Later, the position of $Z_0$ will be defined by requiring to
have the minimal area for the world surface of three-quark system [4]. The
path of integration in Eq. (2) is also arbitrary.

Let the four-points $x$, $y$,$z$ and $x^{\prime }$, $y^{\prime }$,$z^{\prime
}$ be the initial and final positions of three quarks, respectively. Two
particle quantum Green function is defined as [4]:
\begin{equation}
G(xyz,x^{\prime }y^{\prime }z^{\prime })=\langle X_B(x,y,z,C_i)\bar{X}
_B(x^{\prime },y^{\prime },z^{\prime },C_i^{\prime })\rangle ,  \label{(3)}
\end{equation}
were $\bar{X}_B(x^{\prime },y^{\prime },z^{\prime },C_i^{\prime })$
corresponds to the final state of a baryon:
\begin{equation}
\bar{X}_B(x^{\prime},y^{\prime},z^{\prime},C_i^{\prime
})=\epsilon _{mnk}\left[ \bar{q}(x^{\prime})\Phi (x^{\prime
},Z_0^{\prime })\right] _m\left[ \bar{q}(y^{\prime })\Phi
(y^{\prime },Z_0^{\prime })\right] _n\left[ \bar{q}(z^{\prime
})\Phi (z^{\prime},Z_0^{\prime })\right] _k.  \label{(4)}
\end{equation}
Here $\bar{q}=q^{+}\gamma _4$; $q^{+}$ is the Hermitean conjugate quark
field; $\gamma _\mu $ are the Dirac matrices and the brackets $\langle
...\rangle $ mean the path-integrating over gluonic and quark fields:
\begin{equation}
\langle X_B\bar{X}_B\rangle =\int D\bar{q}DqDA_\mu \exp \left\{
iS_{QCD}\right\} X_B\bar{X}_B,  \label{(5)}
\end{equation}
with the $QCD$ action $S_{QCD}$. We imply that the measure $DA_\mu $ in the
path integral (5) includes the well known weight for gluonic fields [11].
The Minkowski space is used here but it is not difficult to go into
Euclidean space to have well defined path-integrals.

Let us introduce the generating functional for Green's function to calculate
the path-integral (5) with respect to quark fields:
\begin{equation}
Z[\bar{\eta },\eta ]=\int D\bar{q}Dq \exp \left\{
iS_{QCD}+i\int dx\left( \bar{q}_a(x)\eta _a(x)+\bar{\eta }
_a(x)q_a(x)\right) \right\},  \label{(6)}
\end{equation}
where we introduce the external colour anticommutative sources $\eta _a$, $
\bar{\eta }_a$. Then the Green function (3) can be written as
\begin{eqnarray*}
G(xyz,x^{\prime }y^{\prime }z^{\prime }) &=&\int DA_\mu \biggl [\epsilon _{abc}
\epsilon_{mnk}\Phi _{aa^{\prime}}(Z_0,x)\Phi _{bb^{\prime }}(Z_0,y)\Phi
_{cc^{\prime}}(Z_0,z) \\
&&\times \Phi _{m^{\prime}m}(x^{\prime },Z_0^{\prime })\Phi
_{n^{\prime}n}(y^{\prime },Z_0^{\prime})\Phi _{k^{\prime
}k}(z^{\prime },Z_0^{\prime})
\end{eqnarray*}
\begin{equation}
\times \delta ^6/\delta \bar{\eta }_{a^{\prime }}(x)\delta \bar{
\eta }_{b^{\prime }}(y)\delta \bar{\eta }_{c^{\prime }}(z)\delta
\eta _{m^{\prime }}(x^{\prime })\delta \eta _{n^{\prime
}}(y^{\prime })\delta \eta _{k^{\prime }}(z^{\prime })Z[\bar{
\eta },\eta ]\biggr ]_{\eta =\bar{\eta }=0}.  \label{(7)}
\end{equation}

In Eq. (7) $Z_0$ and $Z_0^{\prime }$ are the initial and final
positions of the string junction, respectively. The total surface which
consists of world motions of quarks and the path of the string junction must
be minimal. This requirement defines the path of the string junction [4].
Now it is possible to integrate the path-integral in Eq. (7) over quark
fields $\bar{q}$, $q$ as expression (6) is a Gaussian integral. We may
represent the $QCD$ action in the form
\begin{equation}
S_{QCD}=S(A)-\int dx\bar{q}(x)\left( \gamma _\mu D_\mu +m\right) q(x),
\label{(8)}
\end{equation}

where $S(A)$ is an action for gluonic fields with the included ghost fields,
$D_\mu =\partial _\mu -igA_\mu $; $m$ is the quark mass matrix and we imply
the summation on colour and flavour indexes. Inserting Eq. (8) into Eq.
(6) and integrating with respect to quark fields, we arrive at the
expression
\begin{equation}
Z[\bar{\eta },\eta ]=\det (-\gamma _\mu D_\mu -m)\exp
\left\{ iS(A)+i\int dxdy\bar{\eta }(x)S(x,y)\eta (y)\right\} ,
\label{(9)}
\end{equation}
where the classical quark Green function $S(x,y)$ is the solution of the
equation
\begin{equation}
\left( \gamma _\mu D_\mu +m\right) S(x,y)=\delta (x-y).  \label{(10)}
\end{equation}

Using Eq. (9) and calculating the variation derivatives in Eq. (7) we
find the quantum Green function of a baryon:
\[
G(xyz,x^{\prime }y^{\prime }z^{\prime })=\int DA_\mu \det (-\gamma _\mu
D_\mu -m)\exp \left\{ iS(A)\right\} \epsilon _{abc}\epsilon _{mnk}
\]
\[
\times \biggl [S_{bm}^\Phi (y,x^{\prime })S_{an}^\Phi (x,y^{\prime
})S_{ck}^\Phi (z,z^{\prime })-S_{am}^\Phi (x,x^{\prime })S_{bn}^\Phi
(y,y^{\prime })S_{ck}^\Phi (z,z^{\prime })
\]
\begin{equation}
+S_{cm}^\Phi (z,x^{\prime })S_{bn}^\Phi (y,y^{\prime })S_{ak}^\Phi
(x,z^{\prime })-S_{cm}^\Phi (z,x^{\prime })S_{an}^\Phi (x,y^{\prime
})S_{bk}^\Phi (y,z^{\prime })  \label{(11)}
\end{equation}
\[
+S_{am}^\Phi (x,x^{\prime })S_{cn}^\Phi (z,y^{\prime })S_{bk}^\Phi
(y,z^{\prime })-S_{bm}^\Phi (y,x^{\prime })S_{cn}^\Phi (z,y^{\prime
})S_{ak}^\Phi (x,z^{\prime })\biggr ],
\]
where we introduce the following notation for the covariant Green function
\begin{equation}
S_{am}^\Phi (x,x^{\prime })=\Phi _{aa^{\prime }}(Z_0,x)S_{a^{\prime
}m^{^{\prime }}}(x,x^{\prime })\Phi _{m^{\prime }m}(x^{\prime
},Z_0^{\prime }).  \label{(12)}
\end{equation}

The functional determinant in Eq. (9) describes the contribution from the
vacuum polarization and gives the addition quark loops. As a first
approximation we neglect the contribution of the loops. The presence of
different terms in Eq. (11) is connected with the permutations of quark
fields because the quantum Green function being considered. As different terms
in Eq. (11) have the same structure, we consider in detail only one
term. Neglecting the functional determinant we find the approximate
expression for the baryon Green function
\begin{equation}
G_1(xyz,x^{\prime }y^{\prime }z^{\prime })=-\int DA_\mu \exp \left\{
iS(A)\right\} \epsilon _{abc}\epsilon _{mnk}S_{am}^\Phi (x,x^{\prime
})S_{bn}^\Phi (y,y^{\prime })S_{ck}^\Phi (z,z^{\prime }).
\label{(13)}
\end{equation}
Expression (13) is the basic formula for deriving effective action for baryons.

\section {Effective action for baryons}

To calculate some electromagnetic characteristics of nucleons we consider
baryons in external electromagnetic fields. Then covariant derivatives
become $D_\mu =\partial _\mu -igA_\mu -iQA_\mu ^{el}$, where $A_\mu ^{el}$
is the vector-potential of electromagnetic fields, $Q$ is the charge matrix
of quarks, $Q=$diag$(e_1,e_2,...,e_{N_f})$, $e_i$ are charges of quarks, $N_f$
is the number of flavours. Using the proper time
method and Feynman path-integrals we found in [2] the Green function of a
quark which is the solution to Eq. (10):
\[
S(x,x^{\prime })=i\int_0^\infty ds\int_{z(0)=x^{\prime
}}^{z(s)=x}Dz\left( m-\frac i2\gamma _\mu \dot{z}_\mu
(t)\right)
\]
\begin{equation}
\times P_\Sigma \exp \left\{ i\int_0^sdt\left[ \frac 14\dot{z}
_\mu ^2(t)-m^2+e\dot{z}_\mu (t)A_\mu ^{el}(z)+\Sigma _{\mu \nu
}\left( eF_{\mu \nu }^{el}+gF_{\mu \nu }\right) \right] \right\} \Phi
(x,x^{\prime }),  \label{(14)}
\end{equation}
where $F_{\mu \nu }^{el}=\partial _\mu A_\nu ^{el}-\partial _\nu A_\mu ^{el}$,
$F_{\mu \nu }=\partial _\mu A_\nu -\partial _\nu A_\mu -ig\left[ A_\mu,
A_\nu \right] $ are the strength tensors of electromagnetic and gluonic
fields, respectively; $\Sigma _{\mu \nu }=-(i/4)\left[ \gamma _\mu ,\gamma
_\nu \right] $ are the spin matrices; $P_\Sigma $ is the ordering operator
of the spin matrices $\Sigma _{\mu \nu }$; $\dot{z}_\mu
(t)=\partial z_\mu (t)/\partial t;$ $z_\mu (t)$ is the path of the quark
with the boundary conditions $z_\mu (0)=x_\mu ^{\prime }$, $z_\mu
(s)=x_\mu $ and $\Phi (x,x^{\prime })$ is the path ordered product
(see Eq. (2)). Inserting Eq. (14) into Eq. (12) we find
\[
S_{am}^\Phi (x,x^{\prime })=-i\int_0^\infty ds\int_{z(0)=x^{\prime
}}^{z(s)=x}Dz\left( m-\frac i2\gamma _\mu \dot{z}_\mu
(t)\right) P_\Sigma
\]
\begin{equation}
\times \exp \left\{ i\int_0^sdt\left[ \frac 14\dot{z}_\mu
^2(t)-m^2+e\dot{z}_\mu (t)A_\mu ^{el}(z)+\Sigma _{\mu \nu
}(eF_{\mu \nu }^{el}+gF_{\mu \nu})\right] \right\} \left( \Phi _{C_x}(Z_0,Z_0^{\prime
})\right) _{am},  \label{(15)}
\end{equation}
where the contour $C_x$ in Eq. (15) consists of lines between $Z_0$, $x$
and $Z_0^{\prime }$, $x^{\prime }$ and path $z_\mu (t)$. Using the
expression (15) for each quark, the baryon Green function (13) becomes
\[
G_1(xyz,x^{\prime }y^{\prime }z^{\prime })=i\int_0^\infty \prod_j
ds_j\int \prod_j Dz^{(j)}\left( m_j-\frac i2\gamma _\mu
\dot{z}_\mu ^{(j)}(t_j)\right)
\]
\[
\times P_\Sigma \prod_j \exp \left\{ \int_0^{s_j}dt_j\Sigma
_{\mu \nu }\frac \delta {\delta \sigma _{\mu \nu }^{(j)}(t_j)}\right\}
\]
\[
\times \exp \left\{ i\sum_j \int_0^{s_j}dt_j\left[ \frac
14\left( \dot{z}_\mu ^{(j)}(t_j)\right) ^2-m_j^2+e_j
\dot{z}_\mu ^{(j)}(t_j)A_\mu ^{el}(z^{(j)})+e_j\Sigma _{\mu \nu }F_{\mu
\nu }^{el}\right] \right\}
\]
\begin{equation}
\times \langle W(C_xC_yC_z)\rangle,  \label{(16)}
\end{equation}
where $j=1,2,3$; $e_j$ is the charge of the $j$-th quark; the boundary
conditions $z_\mu ^{(1)}(0)=x_\mu ^{\prime }$, $z_\mu ^{(1)}(s_1)=x_\mu$,
$z_\mu ^{(2)}(0)=y_\mu ^{\prime }$, $z_\mu ^{(2)}(s_2)=y_\mu $, $z_\mu
^{(3)}(0)=z_\mu ^{\prime }$, $z_\mu ^{(3)}(s_3)=z_\mu $ are used here and
the Wilson loop is given by (see [4])
\begin{equation}
\langle W(C_xC_yC_z)\rangle =\epsilon _{abc}\epsilon _{mnk}\langle \left(
\Phi _{C_x}(Z_0,Z_0^{^{\prime }})\right) _{am}\left( \Phi
_{C_y}(Z_0,Z_0^{\prime })\right) _{bn}\left( \Phi
_{C_z}(Z_0,Z_0^{\prime })\right) _{ck}\rangle.  \label{(17)}
\end{equation}
We took into account that [4]
\[
\exp \left\{ ig\int_0^sdt\Sigma _{\mu \nu }F_{\mu \nu }\right\} W(C)
=\exp \left\{ \int_0^sdt\Sigma _{\mu \nu }\frac \delta
{\delta \sigma _{\mu \nu }(t)}\right\} W(C),
\]
where $\sigma _{\mu \nu }(t)$ is the surface around the point $z_\mu (t)$.

Contours $C_x$, $C_y$, $C_z$ correspond to three quarks which have paths $
z_\mu ^{(1)}$, $z_\mu ^{(2)}$, $z_\mu ^{(3)}$ and masses $m_1$, $m_2$, $m_3$,
respectively. Relationship (16) is the generalization of one [3,4] for
the case of quarks placed in external electromagnetic fields which
possess spins.

We imply further that the average distance between quarks $\langle r\rangle $ is
greater than the time fluctuations (in units $c=\hbar =1$) of the gluonic
fields $T_g$: $\langle r\rangle >T_g$. As $T_g\simeq 0.2\div 0.3$ fm [12,13]
so the condition $\langle r\rangle >T_g$ is valid not only for asymptotic
baryon states of the Regge trajectories with large angular momenta of the
baryon but also for lower baryon states. The asymptotic of the average
Wilson loop integral obeys then the area law and is given by (in the
Minkowski space):
\begin{equation}
\langle W(C_xC_yC_z)\rangle =\exp \left\{ -i\sigma \left( S_1+S_2+S_3\right)
\right\},  \label{(18)}
\end{equation}
where $\sigma$ is the string tension and $S_j$ ($j=1,2,3$) is the minimal
surface bounded by the trajectories of the quark $q_j$ and string junction $
Z_0$. The path of the string junction is defined by the requirement that the
sum $S_1+S_2+S_3$ is minimal [3,4]. Following [3,4], new variables are
introduced:
\begin{equation}
\tau =\frac{t_1T}{s_1}=\frac{t_2T}{s_2}=\frac{t_3T}{s_3},\hspace{0.3in}\mu
_j=\frac T{2s_j},  \label{(19)}
\end{equation}
where $\tau $ means the proper time for every quark and $\mu _j$ ($j=1,2,3$
) is the dynamical mass of the $j-$th quark. Using Eqs. (18), (19) from
Eq. (16) we arrive at
\[
G_1(xyz,x^{\prime }y^{\prime }z^{\prime })=-i\frac{T^3}8\int_0^\infty
\prod_j \frac{d\mu _j}{\mu _j^2}\int \prod_j
Dz^{(j)}\left( m_j-i\mu _j\gamma _\mu \dot{z}_\mu ^{(j)}(\tau
)\right) \times
\]
\[
\times P_\Sigma \prod_j \exp \left\{ \frac 1{2\mu _j}\Sigma
_{\mu \nu }\int_0^Td\tau \frac \delta {\delta \sigma _{\mu \nu }^{(j)}(\tau
)}\right\} \times
\]
\[
\times \exp \{i\int_0^Td\tau \sum_j \left[ \frac 12\mu
_j\left( \dot{z}_\mu ^{(j)}(\tau )\right) ^2-\frac{m_j^2}{2\mu
_j}+e_j\dot{z}_\mu ^{(j)}(\tau )A_\mu ^{el}(z^{(j)})+\frac{e_j}{
2\mu _j}\Sigma _{\mu \nu }F_{\mu \nu }^{el}\right]
\]
\begin{equation}
-i\sigma \left( S_1+S_2+S_3\right) \}.  \label{(20)}
\end{equation}

The integral in the last exponential factor of Eq. (20) represents the
effective action for three-quark system (baryon) taking into account the
spins of quarks. So $T$ is the time of the observation, $\tau $ is the
proper time of quarks; $m_j$ and $\mu _j$ are the current and dynamical
masses of $j-$th quark. It follows from Eq. (20) that there is
integration over dynamical masses $\mu _j$ to get the Green function of the
three-quark system. Pre-exponential factors in Eq. (20) allow us to calculate
in principle the spin-spin and spin-orbit contributions to the effective
action. As a first approximation (see [3,4]) we neglect the short-range spin
corrections and consider therefore scalar quarks. The terms $e_j\Sigma _{\mu
\nu }F_{\mu \nu }^{el}$ which describe the interaction of the magnetic field
with the spins of quarks will be omitted. With this assumption we arrive at the
effective action for baryons in external electromagnetic fields
\begin{equation}
B=\int_0^Td\tau \sum_j \left[ \frac 12\mu _j\left(
\dot{z}_\mu ^{(j)}(\tau )\right) ^2-\frac{m_j^2}{2\mu _j}+e_j
\dot{z}_\mu ^{(j)}(\tau )A_\mu ^{el}(z^{(j)})\right] -\sigma \left(
S_1+S_2+S_3\right).  \label{(21)}
\end{equation}

The case when electromagnetic fields are absent was considered in [3,4]. The
terms which describe the interaction of quarks with electromagnetic fields
are essential for us because we are going to calculate electromagnetic
characteristics of nucleons. It is convenient to introduce new variables $
R_\mu $, $\xi _\mu $ and $\eta _\mu $ instead of $z_\mu ^{(j)}$ in
accordance with relationships [3,4]:
\[
z_\mu ^{(1)}=R_\mu +\left( \frac{\mu \mu _3}{M\left( \mu _1+\mu _2\right) }
\right) ^{1/2}\xi _\mu -\left( \frac{\mu \mu _2}{\mu _1\left( \mu _1+\mu
_2\right) }\right) ^{1/2}\eta _\mu,
\]
\[
z_\mu ^{(2)}=R_\mu +\left( \frac{\mu \mu _3}{M\left( \mu _1+\mu _2\right) }
\right) ^{1/2}\xi _\mu +\left( \frac{\mu \mu _1}{\mu _2\left( \mu _1+\mu
_2\right) }\right) ^{1/2}\eta _\mu,
\]
\begin{equation}
z_\mu ^{(3)}=R_\mu -\left( \frac{\mu \left( \mu _1+\mu _2\right) }{M\mu _3}
\right) ^{1/2}\xi _\mu,  \label{(22)}
\end{equation}
where $R_\mu $ is the center of mass coordinate of a baryon; $\xi _\mu $ and
$\eta _\mu $ are relative coordinates of quarks, $M=\mu _1+\mu _2+\mu _3$ is
the sum of dynamical masses of quarks. The arbitrary mass parameter $\mu $
in Eq. (22) defines the scale of relative coordinates $\xi _\mu $ and $
\eta _\mu $. After the substitution (22), the measure $\prod_j
Dz^{(j)}$ transforms into $DRD\eta D\xi $ in path-integral (20).

Let us consider the uniform and constant external electromagnetic fields.
Then the vector-potential of electromagnetic fields can be represented as
\begin{equation}
A_v^{el}(z^{(j)})=\frac 12F_{\mu \nu }^{el}z_\mu ^{(j)}.  \label{(23)}
\end{equation}

Inserting Eqs. (22), (23) into Eq. (21) we find the effective action
for the three quark system (see [5]) in the form
\begin{equation}
B=\int_0^Td\tau \left[ \frac M2\dot{R}_\nu^2+\frac \mu 2\left(
\dot{\xi }_\nu ^2+\dot{\eta }_\nu ^2\right) -
\sum_j \frac{m_j^2}{2\mu _j}\right] -\sigma \left(
S_1+S_2+S_3\right) +\Delta B,  \label{(24)}
\end{equation}
\[
\Delta B=\frac 12F_{\nu \mu }^{el}\int_0^Td\tau \biggl [e\dot{R}
_\mu R_\nu +\dot{\lambda \xi _\mu }\xi _\nu +\rho
\dot{\eta _\mu }\eta _\nu
\]
\begin{equation}
+\gamma \left( \dot{R_\mu }\xi _\nu +\dot{\xi }
_\mu R_\nu \right) +\delta \left( \dot{R_\mu }\eta _\nu +
\dot{\eta }_\mu R_\nu \right) +\delta \sqrt{\frac{\mu \mu _3}{
\left( \mu _1+\mu _2\right) M}}\left( \dot{\xi }_\mu \eta _\nu
+ \dot{\eta }_\mu \xi _\nu \right) \biggr ],  \label{(25)}
\end{equation}
where we introduce parameters:
\[
\gamma =\sqrt{\frac \mu M}\left[ \left( e_1+e_2\right) \sqrt{\frac{\mu _3}{
\mu _1+\mu _2}}-e_3\sqrt{\frac{\mu _1+\mu _2}{\mu _3}}\right],
\]
\[
\delta =\sqrt{\frac \mu {\mu _1+\mu _2}}\left( e_2\sqrt{\frac{\mu _1}{\mu _2}
}-e_1\sqrt{\frac{\mu _2}{\mu _1}}\right),
\]
\[
\rho =\frac \mu {\mu _1+\mu _2}\left( \frac{e_1\mu _2}{\mu _1}+\frac{e_2\mu
_1}{\mu _2}\right),
\]
\begin{equation}
\lambda =\frac \mu M\left[ \frac{\left( e_1+e_2\right) \mu _3}{\mu _1+\mu _2}
+\frac{e_3\left( \mu _1+\mu _2\right) }{\mu _3}\right],  \label{(26)}
\end{equation}
and $e=e_1+e_2+e_3$ is the charge of a baryon. As a particular case, when
electromagnetic fields are absent ($\Delta B=0$) we arrive at the action
derived in [3]. It follows from Eq. (24) that the center of mass
coordinate $R_\mu $ is separated from relative coordinates $\xi _\mu $ and $
\eta _\mu $ and $\mu $ plays the role of the mass of the $\xi _\mu $, $\eta
_\mu $ excitations. Following [3,4], the straight line approximation for
strings and the asymmetric quark-diquark structure of baryons will be
assumed. The asymmetric configuration (see also [9,10]) means that two
quarks $q^{(1)}$ and $q^{(2)}$ are near each other and quark $q^{(3)}$ is
farther from them. This case is preferable [9,10] and the slope of
linear baryon Regge trajectories is the same as for mesons [3,4]. Then $\mid
{\bf \xi }\mid \gg \mid {\bf \eta }\mid $ ($\mid {\bf \xi }\mid =
\sqrt{\xi _1^2+\xi _2^2+\xi _3^2}$) and the coordinate $\eta _\mu $ can be
ignored. We neglect therefore the surfaces $S_1$, $S_2$ and assume for $S_3$
the following expression [3,4]:
\begin{equation}
S_3=b\int_0^Td\tau \mid {\bf \xi }\mid,  \label{(27)}
\end{equation}
where
\begin{equation}
b=\sqrt{\frac{\mu \left( \mu _1+\mu _2\right) }{M\mu _3}}+\sqrt{\frac{\mu
\mu _3}{M\left( \mu _1+\mu _2\right) }}.  \label{(28)}
\end{equation}

Equation (27) takes into account the confinement of quarks and gives the
linear potential between quarks. Using the definition $B=\int_0^Td\tau
{\cal L}$, where ${\cal L}$ is the Lagrangian, from Eq. (25) we
arrive at the effective Lagrangian for baryons
\begin{equation}
{\cal L}_{eff}=\frac M2\dot{R}_\nu^2+\frac \mu 2
\dot{\xi }_\nu ^2-\sum_j \frac{m_j^2}{2\mu _j}-\sigma b\mid
{\bf \xi }\mid +{\cal L}^{el}.  \label{(29)}
\end{equation}

Here we neglect the coordinate $\eta _\mu $ and introduce the notation:
\begin{equation}
{\cal L}^{el}=\frac 12F_{\nu \mu }^{el}\left[ e\dot{R}_\mu
R_\nu +\lambda \dot{\xi }_\mu \xi _\nu +\gamma \left(
\dot{R_\mu }\xi _\nu +\dot{\xi }_\mu R_\nu \right) \right].
\label{(30)}
\end{equation}
Lagrangian ${\cal L}^{el}$ describes the electromagnetic interaction of
the string. It follows from Eq. (3.22) that
\begin{equation}
z_\mu ^{(3)}-\frac 12\left( z_\mu ^{(1)}+z_\mu ^{(2)}\right) =-b\xi _\mu
+\frac 12\sqrt{\frac \mu {\mu _1+\mu _2}}\left( \frac{\mu _2-\mu _1}{\sqrt{
\mu _1\mu _2}}\right) \eta _\mu.  \label{(31)}
\end{equation}

As the second term in Eq. (31) is small, the coordinate $\xi _\mu $ is
proportional to the ``distance'' between quark $q^{(3)}$ and the center of
mass of quarks $q^{(1)}$ and $q^{(2)}$ which form a diquark. At the large
time $T$ limit $\xi _4=0$, $R_4=i\tau $ [3,4] and Lagrangian (30) takes
the form
\begin{equation}
{\cal L}^{el}=e\left( {\bf RE}\right) +\gamma \left( {\bf \xi E}
\right) -\frac 12\epsilon _{mnk}H_k\left( e\dot{R}_mR_n+\lambda
\dot{\xi }_m\xi _n+\gamma \dot{(R_m}\xi _n+
\dot{\xi }_mR_n)\right),  \label{(32)}
\end{equation}
where the electric field $E_k=iF_{k4}$ and magnetic field $H_k=(1/2)\epsilon
_{kmn}F_{mn}$. To clarify the physical meaning of the terms in Eq. (32),
let us consider the dipole moment of quarks. Using the definition of the
electric dipole moment and Eq. (22) we have
\begin{equation}
{\bf d}=\sum_j e_j{\bf z}^{(j)}=e{\bf R}+\gamma
{\bf \xi }+\delta {\bf \eta }.  \label{(33)}
\end{equation}

So, first two terms in Eq. (32) ( neglecting the coordinate $\eta _\mu $
) describe the interaction of the dipole moment of quarks with the electric
field in accordance with the expression for the potential energy: $U=-(
{\bf dE})$. The magnetic moment is given by
\[
m_k=\frac 12\epsilon _{mnk}\sum_j e_jz_m^{(j)}\dot
{z}_n^{(j)}=\frac 12\epsilon _{mnk}\biggl [eR_m\dot{R}
_n+\lambda \xi _m\dot{\xi }_n+\gamma \left( R_m\dot
{\xi }_n+\xi _m\dot{R}_n\right)
\]
\begin{equation}
+\rho \eta _m\dot{\eta }_n+\delta \left( R_m\dot{
\eta }_n+\eta _m\dot{R}_n\right) +\delta \sqrt{\frac{\mu \mu _3
}{\left( \mu _1+\mu _2\right) M}}\left( \xi _m\dot{\eta }
_n+\eta _m\dot{\xi }_n\right) \biggr ].  \label{(34)}
\end{equation}

It follows from Eqs. (25), (32) that there is an interaction of the
magnetic field with the magnetic moment in such a way that the interaction
energy is $U=-\left( {\bf mH}\right)$. So Lagrangian (32) describes
the interaction of electric and magnetic moments of baryons with electric
and magnetic fields, respectively.

\section {Mean relative coordinate and electric polarizabilities of nucleons}

Now we estimate the mean size of baryons.
At the large time $T$ limit $\stackrel{\cdot }{\xi }_0=0$ ($\xi _4=i\xi _0)$,
$R_4=i\tau $ [3,4] and therefore only three dimensional quantities $R_k$
and $\xi _k$ are important. From Eq. (32) we find three momenta
corresponding to the center of mass coordinate $R_k$ and relative coordinate
$\xi _k:$
\[
\Pi _k=\frac{\partial {\cal L}_{eff}}{\partial \dot{R}_k}=M
\dot{R}_k-\frac 12\epsilon _{knm}H_m\left( eR_n+2\gamma \xi
_n\right),
\]
\begin{equation}
\pi _k=\frac{\partial {\cal L}_{eff}}{\partial \dot{\xi }_k}
=\mu \dot{\xi }_k-\frac 12\epsilon _{knm}H_m\left( \lambda \xi
_n+2\gamma R_n\right),  \label{(35)}
\end{equation}

Here we take into account that Lagrangian is defined within the accuracy of
the total derivative on time. Then the effective Hamiltonian ${\cal H}
_{eff}=\Pi _k\dot{R}_k+\pi _k\dot{\xi }_k-{\cal L}_{eff}$ corresponding to
the quark-diquark structure of a baryon is given by
\begin{equation}
{\cal H}_{eff}=\sum_j \frac{m_j^2}{2\mu _j}+\frac M2+\frac
M2\dot{R}_k^2+\frac \mu 2\dot{\xi }_k^2+\sigma
b\mid {\bf \xi }\mid -e\left( {\bf ER}\right) -\gamma \left( {\bf
E\xi }\right),  \label{(36)}
\end{equation}

The Hamiltonian for baryons Eq. (36) looks like the one for mesons [2] because we
consider basically the string between quark and diquark. Therefore the
calculations of mean coordinates and electromagnetic polarizabilities is
the same. But in the case of baryons there are more parameters and the
analysis is more complicated.

With the help of Eq. (35) the effective Hamiltonian (4.36) is rewritten as
\[
{\cal H}_{eff}=\sum_j \frac{m_j^2}{2\mu _j}+\frac M2+\frac
1{2M}\left[ {\bf \Pi} +\frac e2\left( {\bf R}\times {\bf H}\right)
+\gamma \left( {\bf \xi }\times {\bf H}\right) \right] ^2
\]
\begin{equation}
+\frac 1{2\mu }\left[ {\bf \pi }+\frac \lambda 2\left( {\bf \xi }
\times {\bf H}\right) +\gamma \left( {\bf R}\times {\bf H}\right)
\right] ^2+\sigma b\mid {\bf \xi }\mid -e\left( {\bf ER}\right)
-\gamma \left( {\bf E\xi }\right),  \label{(37)}
\end{equation}
where $\left( {\bf \xi }\times {\bf H}\right) _k=\epsilon _{mnk}\xi
_mH_n$. The mass of a baryon ${\cal M}(\mu _j)$ is defined here as a
solution to equation
\begin{equation}
{\cal H}_{eff}\Phi ={\cal M}(\mu _j)\Phi.  \label{(38)}
\end{equation}

In according to the Noether theorem, the momentum ${\bf \Pi }$ is
conserved and we can put ${\bf R}={\bf \Pi }=0$ in Eq. (37). To find the
solution to Eq. (38) we can use the substitution $\pi _k=-i\partial
/\partial \xi _k$. Then the mass of a baryon ${\cal M}(\mu _j)$ is given
by
\begin{equation}
{\cal M}(\mu _j)=\sum_j \frac{m_j^2}{2\mu _j}+\frac
M2+\epsilon (\mu _j,{\bf E,H}),  \label{(39)}
\end{equation}
where $\epsilon (\mu _j,{\bf E,H})$ is the eigenvalue of the equation
\[
\left\{ \frac 1{2\mu }\left[ -i\frac {\partial} {\partial {\bf \xi
}}-\frac \lambda 2\left( {\bf \xi }\times {\bf H}\right) \right] ^2+
\frac{\gamma ^2}{2M}\left( {\bf \xi }\times {\bf H}\right) ^2+\sigma
b\mid {\bf \xi }\mid -\gamma \left( {\bf E\xi }\right) \right\} \Phi
\]
\begin{equation}
=\epsilon (\mu _j,{\bf E,H})\Phi.  \label{(40)}
\end{equation}

The term $(\gamma ^2/(2M))\left( {\bf \xi }\times {\bf H}\right) ^2$
in Eq. (40) is due to the recoil of the string. In nonrelativistic models
the effect of the recoil was studied in [13] (see also [14]). As we
neglected the spin of baryons here, there is no interaction of spin with the
external magnetic field. It is not difficult to take into account such
interaction. Eq. (40) is like the equation for mesons [2] and therefore we
write out the solution for the ground state. If ${\bf E}={\bf H}=0$,
the solution to Eq. (40) when the orbital quantum number $l=0$ is given by
\begin{equation}
\Phi (\rho,\theta,\phi)=\frac N\rho Ai(\rho -a(n))Y_{lm}(\theta,\phi),
\label{(41)}
\end{equation}
where $\rho =\sqrt{\rho _1^2+\rho _2^2+\rho _3^2}$, $\rho _k=(2\mu b\sigma
)^{1/3}\xi _k$, $Ai(\rho -a(n))$ is the Airy function, $Y_{lm}(\theta,\phi)$
is the spherical function, $N$ is the normalization constant and $a(n)$ are
the Airy function zeros so that $a(1)=2.3381$, $a(2)=4.0879$, $a(3)=5.52$
and so on [15]. For the ground state of baryons the principal quantum number
$n=1$. For the excited states it is necessary to choose the corresponding
value of $n=n_r+l+1$, where $n_r$ is the radial quantum number. The
eigenvalue of Eq. (40) (at ${\bf E}={\bf H}=0$) is
\begin{equation}
\epsilon (\mu _j)=\left( 2\mu \right) ^{-1/3}\left( b\sigma \right)
^{2/3}a(n)=(\sigma )^{2/3}a(n)\left[ \frac M{2\mu _3\left( \mu _1+\mu
_2\right) }\right] ^{1/3}.  \label{(42)}
\end{equation}
The condition of the minimum of the baryon mass (39) ($\partial {\cal M}
(\mu _j)/\mu _j=0$) at $m_1=m_2$ (and $\mu _1=\mu _2$), with the help of Eq.
(41) gives the dynamical mass of a diquark:
\begin{equation}
\mu _3^{(0)}=\mu _1^{(0)}+\mu _2^{(0)}=\sqrt{\sigma }\left[ \frac{a(n)}
3\right] ^{3/4}.  \label{(43)}
\end{equation}

This value is different from one [3] obtained for large angular momentum.
Using Eq. (43) we arrive from Eq. (39) at the expression for the mass of
a baryon (see also [3]):
\begin{equation}
{\cal M}(\mu _j)=\frac{m_3^2+4m_1^2}{2\mu _3^{(0)}}+4\mu _3^{(0)}.
\label{(44)}
\end{equation}

To estimate the baryon mass, the value of the string tension $\sigma =0.15$
GeV$^2$ will be used [3,4]. Neglecting the small current masses of quarks $
m_j$ we find from Eqs. (43), (44) the mass of a diquark for $n=1$: $\mu
_3^{(0)}=320$ MeV and the nucleon mass : ${\cal M}(\mu _j)=1.28$ GeV [5].
This value of the nucleon mass is a little greater then real nucleon mass
because spin-spin and spin-orbit forces were omitted.

Now we consider the mean relative coordinates of nucleons on the basis of the
virial theorem which gives (as for the case of mesons [2]) the mean potential
energy $\langle U\rangle =2\langle T\rangle $, where $\langle T\rangle $ is
the mean kinetic energy. Then using the relation $\langle T\rangle +\langle
U\rangle =\epsilon (\mu _j)$, we arrive at
\begin{equation}
\langle U\rangle =\frac 23\epsilon (\mu _j)=\frac 23(2\mu )^{-1/3}(b\sigma
)^{2/3}a(n).  \label{(45)}
\end{equation}

Comparing Eq. (45) with the relation $\langle U\rangle =b\sigma \langle
\sqrt{{\bf \xi }^2}\rangle $ gives the following expression
\begin{equation}
\langle \sqrt{{\bf \xi }^2}\rangle =\frac 23(2\mu b\sigma
)^{-1/3}a(n)=\sigma ^{-1/4}\sqrt{\frac 2\mu }\left( \frac{a(n)}3\right)
^{9/8}.  \label{(46)}
\end{equation}

In accordance with Eq. (31) the size of the nucleon is characterized by
the value $\mid {\bf z}^{(3)}-(1/2)({\bf z}^{(1)}+{\bf z}
^{(2)})\mid \simeq \mid b{\bf \xi }\mid $. Introducing the notation $
{\bf r=}b{\bf \xi }$, from Eq. (46) we have
\begin{equation}
\langle \sqrt{{\bf r}^2}\rangle =\frac 2{\sqrt{\sigma }}\left( \frac{a(n)}
3\right) ^{3/4}.  \label{(47)}
\end{equation}

The same expression was found in [1,2] for mesons. For the quark-diquark system
the string tension coincides with those of mesons and therefore the
quark-diquark system has approximately the same size as mesons. Using $
\sigma =0.15$ GeV$^2$ and $a(1)=2.2281$ we find the mean size of the
nucleons
\begin{equation}
\langle \sqrt{{\bf r}^2}\rangle =0.84~fm.  \label{(48)}
\end{equation}

The experimental value of the charge radii of the proton and
neutron are $ \sqrt{\langle r_p^2\rangle }=0.86$ fm [16], $\langle
r_n^2\rangle =-0.113\pm 0.005~\mbox{fm}^2$ [17]. In accordance
with Eq. (4.43) the center of mass of the quark-diquark system is
situated in the center between quark $q^{(3)}$ and diquark $\left(
q^{(1)},q^{(2)}\right) $ and the mean radius of a nucleon is
$\left( 1/2\right) \langle \sqrt{{\bf r}^2}\rangle =0.42$ fm which
is the reasonable value. It should be noted that charge radii of
hadrons are defined from electromagnetic formfactors.

Now let us consider the case when ${\bf H}=0$, ${\bf E}\neq 0.$ It is
possible to assume as an approximation that ${\bf E}\parallel {\bf \xi }$ [5],
i.e. external electric field is parallel to the string which connects quark $
q^{(3)}$ with diquark $\left( q^{(1)},q^{(2)}\right) $. So we neglect the
rotation of the string. It is justified only for the ground state when the
orbital quantum number $l=0$. Introducing the effective string tension
\begin{equation}
\sigma _{eff}=\sigma -\frac \gamma bE,  \label{(49)}
\end{equation}
we arrive from Eq. (40) at the eigenvalue $\epsilon (\mu _j,{\bf E}
)=\left( 2\mu \right) ^{-1/3}\left( b\sigma _{eff}\right) ^{2/3}a(n)$. From
(43), (44), by neglecting the small terms containing the current masses
we find the mass of a baryon
\begin{equation}
{\cal M}(\mu _j,{\bf E})=4\sqrt{\sigma _{eff}}\left[ \frac{a(n)}
3\right] ^{3/4},  \label{(50)}
\end{equation}
which depends on the external electric field. Inserting Eq. (49) into Eq.
(50) and expanding it in a small electric field one yields
\begin{equation}
{\cal M}(\mu _j,{\bf E})\simeq \left[ \frac{a(n)}3\right] ^{3/4}\left(
4\sqrt{\sigma }-\frac q{\sqrt{\sigma }}E-\frac{q^2}8\sigma ^{-3/2}E^2\right),
\label{(51)}
\end{equation}
where Eq. (43) was used and $q=e_1+e_2-e_3$. We write here only terms of
the expansion ${\cal M}(\mu _j,{\bf E})$ in small electric field up to $
E^2$. The first term in Eq. (51) gives the mass of a baryon. The second one is
connected with the potential energy of a dipole moment of quarks in the
external electric field $U=-{\bf dE}$. From Eq. (33) when the center of
mass coordinate ${\bf R}=0$ and $\mid {\bf \xi }\mid $ $\gg \mid
{\bf \eta }\mid $, the dipole moment of quark-diquark system is ${\bf
d }\simeq \gamma {\bf \xi }=(\gamma /b){\bf r}$. Comparing the
potential energy of a dipole $U=-(\gamma /b)rE$ (at ${\bf E\parallel r}$)
with the second term of Eq. (51): $-(q/\sqrt{\sigma })[a(n)/3]^{3/4}E$ we
arrive at the expression for the mean relative coordinate $r=(2/\sqrt{\sigma
} )[a(n)/3]^{3/4}$ which coincides with Eq. (47). Here we define the
electric dipole moment of quark-diquark system more precisely as compared
with the letter [5] and as a result the mean relative coordinate of a baryon
Eq. (47) coincides with that for a meson. The third term in Eq. (51)
describes the potential energy due to the electric polarizability of a
baryon. The polarization potential is given by (see [18,14])
\begin{equation}
U(\alpha ,\beta )=-\frac 12\alpha E^2-\frac 12\beta H^2,  \label{(52)}
\end{equation}
where $\alpha $, $\beta $ are electric and magnetic static polarizabilities
of hadrons, respectively. From Eq. (51) by comparing the quadratic term in
${\bf E}$ with Eq. (52) we arrive at the electric polarizability of a
baryon:
\begin{equation}
\alpha =\frac{q^2}4\left[ \frac{a(n)}3\right] ^{3/4}\sigma ^{-3/2}.
\label{(53)}
\end{equation}

Let us consider the estimation of the electric polarizability for the proton
$p=uud$. There are two possibilities for a proton as a quark-diquark system:
a) the quark $q^{(3)}=d$ and diquark $(q^{(1)}q^{(2)})=(uu)$, so the
electric charges $e_1=e_2=(2e)/3$, $e_3=-e/3$ and parameter $
q=e_1+e_2-e_3=(5e)/3$; b) the quark $q^{(3)}=u$, diquark $
(q^{(1)}q^{(2)})=(ud)$ and the electric charges $e_1=e_3=(2e)/3$, $e_2=-e/3$
and parameter $q=e_1+e_2-e_3=-e/3$. It should be noted that there are no
permutations of quarks here which occur in the Green function Eq. (11).
Using the value of the string tension $\sigma =0.15$
GeV$^2$ and $a(1)=2.3381$ from Eq. (53) we find the static polarizability
of a proton in Gaussian units for two cases
\begin{equation}
a)~ \alpha _p=5.56\times 10^{-4}~ fm^3\;\hspace{0.3in}
b)~ \alpha _p=0.22\times 10^{-4}~ fm^3.  \label{(54)}
\end{equation}

The generalized electric polarizability which is extracted from measurements
of the Compton scattering cross sections is given by [14]
\begin{equation}
\overline{\alpha }=\alpha +\Delta \alpha,  \label{(55)}
\end{equation}
\begin{equation}
\Delta \alpha =\frac{e^2r_E^2}{3M_N}+\frac{e^2\left( \kappa^2
+1\right) }{4M_N^3}, \label{(56)}
\end{equation}
where $M_N$ is the mass of a nucleon, $r_E$ is electric radius and magnetic
moment of the hadron $\mu =(e\kappa )/(2M_N)$. Using the experimental values
of the electric radius and magnetic moment of a proton we have $\Delta \alpha
_p=(4.5\pm 0.1)\times 10^{-4}$ fm$^3$ [14]. From (54) the total electric
polarizability of a proton is becomes
\begin{equation}
a)~ \bar{\alpha }_p=10\times 10^{-4}~  fm^3\;
\hspace{0.3in}b)~ \bar{\alpha }_p=4.7\times 10^{-4}~ fm^3.
\label{(57)}
\end{equation}

The configuration of a proton a), when the quark $q^{(3)}=d$ and
diquark $ (q^{(1)}q^{(2)})=(uu)$ is more favorable as the
experimental values of electric polarizability are
\[
\bar{\alpha }_p^{\exp }=\left( 12.1\pm 0.8\pm 0.5\right) \times
10^{-4}~ fm^3~~~ [19],
\]
\[
\bar{\alpha }_p^{\exp }=\left( 10.6\pm 1.2\pm 1.0\right) \times
10^{-4}~ fm^3~~~ [20],
\]
\[
\bar{\alpha }_p^{\exp }=(9.8\pm 0.4\pm 1.1)\times 10^{-4}~ fm^3~~~
[21].
\]
So in the case a) we have a good agreement with experimental data.

For the neutron $n=udd$ there are also two possibilities: a) the quark $
q^{(3)}=u$ and diquark $(q^{(1)}q^{(2)})=(dd)$, so the electric charges $
e_1=e_2=-e/3$, $e_3=(2e)/3$ and parameter $q=e_1+e_2-e_3=-(4e)/3$; b) the
quark $q^{(3)}=d$, diquark $(q^{(1)}q^{(2)})=(ud)$ and the electric charges $
e_3=e_2=-e/3$, $e_1=(2e)/3$ and parameter $q=e_1+e_2-e_3=(2e)/3$. Inserting
these parameters into Eq. (53) one gives
\begin{equation}
a)~ \alpha _n=3.56\times 10^{-4}~ fm^3\;\hspace{0.3in}
b)~ \alpha _n=0.89\times 10^{-4}~ fm^3.  \label{(58)}
\end{equation}

For the neutron $\Delta \alpha _n=0.62\times 10^{-4}$ fm$^3$ [14] and the
generalized electric polarizability of a neutron is given by
\begin{equation}
a)~ \bar{\alpha }_n=4.2\times 10^{-4}~ fm^3\;
\hspace{0.3in}b)~ \bar{\alpha }_n=1.5\times 10^{-4}~ fm^3.
\label{(59)}
\end{equation}

The experimental situation for a neutron is more complicated as
there are different experimental data:
\[
\bar{\alpha }_n^{\exp }=(12.6\pm 1.5\pm 2.0)\times 10^{-4}~
fm^3~~~ [22],
\]
\[
\alpha_n =(8.8\pm 2.4\pm 3)\times 10^{-4}~ fm^3~~~ [23].
\]
 The recent data [23] are close to
 case a) in Eqs. (59) with quark $q^{(3)}=u$
and diquark $(q^{(1)}q^{(2)})=(dd)$. The value of $\bar{\alpha }
_n$ in the situation a) in Eq. (59) is close to one obtained in
the oscillator nonrelativistic quark model [24-26,18,14]. The
results of calculations in the framework of the dispersion sum
rule [27] and the chiral perturbation theory (CHPT) [28] are
closer to the case a).

\section {Diamagnetic polarizabilities of nucleons}

In according to Eq. (52) for calculating the magnetic polarizability of
nucleons one needs to compare it with Eq. (37). The effective Hamiltonian
(37) (at ${\bf E}=0$, ${\bf R}={\bf \Pi }=0$) can be cast into
\[
{\cal H}_{eff}={\cal H}_0+{\cal H}_{int},
\]
\[
{\cal H}_0=\sum_j \frac{m_j^2}{2\mu _j}+\frac M2-\frac
1{2\mu }\frac{\partial ^2}{\partial \xi _j^2}+\sigma b\mid {\bf \xi }\mid,
\]
\begin{equation}
{\cal H}_{int}=-\frac \lambda {2\mu }{\bf HL}+\left( \frac{\lambda ^2}{
8\mu }+\frac{\gamma ^2}{2M}\right) \left[ \left( {\bf \xi }\times {\bf
 H}\right) \right] ^2,  \label{(60)}
\end{equation}
where $L_k=-i\epsilon _{kmn}\xi _m\partial _n$ is the angular momentum and $
\partial _n$ =$\partial /\partial \xi _n$. Without loss of generality we can
choose the direction of the magnetic field on the third axis, i.e. ${\bf
H }=(0,0,H).$ Then the Hamiltonian of an interaction of quarks with the
magnetic field is given by
\begin{equation}
{\cal H}_{int}=-\frac{\lambda H}{2\mu }L_3+H^2\left( \frac{\lambda ^2}{
8\mu }+\frac{\gamma ^2}{2M}\right) \left( \xi _1^2+\xi _2^2\right),
\label{(61)}
\end{equation}
where $L_3=i\left( \xi _2\partial _1-\xi _1\partial _2\right)$. Considering
the small external magnetic field, the perturbative theory can be applied.
Using the perturbative method [28] one arrives at the shift of the energy
\[
\Delta {\cal E}_n=\langle n\mid \left[ -\frac{\lambda H}{2\mu }L_3+\left(
\frac{\lambda ^2}{8\mu }+\frac{\gamma ^2}{2M}\right) H^2\xi ^2\sin \vartheta
\right] \mid n\rangle
\]
\begin{equation}
+\sum_{n^{\prime}}  \frac{\mid \langle n^{\prime }\mid -\left(
\lambda HL_3\right) /\left( 2\mu \right) \mid n\rangle\mid ^2 }{{\cal E}_n-
{\cal E}_{n^{\prime }}},  \label{(62)}
\end{equation}
where $\vartheta $ is the angle between coordinate ${\bf \xi }$ and
magnetic field ${\bf H}$. If the first term in Eq. (62) is not equal to zero then
the second and third terms are smaller, and the main contribution to the energy
comes from the first term. This occurs when the orbital quantum number $l>0$. In
the case $l=0$, the shift of the energy due to the interaction with the magnetic
field is defined by the second term in Eq. (62).

After averaging Eq. (62) and taking into
account the equation $\left( 1/4\pi \right) \int \sin ^2\vartheta d\Omega
=2/3$ we find for the ground state when $l=0$, $L_3\mid 0\rangle =0$, the
shift of energy
\begin{equation}
\Delta {\cal E}_n=\left( \frac{\lambda ^2}{4\mu }+\frac{\gamma ^2}
M\right) \frac{H^2}3\langle {\bf \xi }^2\rangle,  \label{(63)}
\end{equation}
where $\langle {\bf \xi }^2\rangle=\langle 0\mid {\bf \xi}^2\mid 0\rangle$, and
$\mid 0\rangle$ means the wave function of the ground $s$-state. We took into account
that the first and the third terms in Eq. (62) equal zero because $L_3\mid 0\rangle=0$.
Here we ignore the spin interactions of a baryon with the external magnetic field
and therefore only diamagnetic polarizability can be defined from Eq.
(63). Using the definition of the relative coordinate ${\bf r}=b{\bf
 \xi }$ and comparing Eq. (63) with Eq. (52) gives the diamagnetic
polarizability of a baryon
\begin{equation}
\beta ^{dia}=-\left( \frac{\lambda ^2}{4\mu }+\frac{\gamma ^2}M\right) \frac
2{3b^2}\langle {\bf r}^2\rangle.  \label{(64)}
\end{equation}

As required, the diamagnetic polarizability is negative and Eq. (64)
is like the Langevin formula for the magnetic susceptibility of atoms. The
similar expression was derived in [2] for mesons. Take into account Eqs.
(26), (43), expression (64) is rewritten as
\begin{equation}
\beta ^{dia}=-\left( \frac{e^2}4+q^2\right) \frac{\langle {\bf r}
^2\rangle }{6M},  \label{(65)}
\end{equation}
where $M=\mu _1^{(0)}+\mu _2^{(0)}+\mu _3^{(0)}=2\mu _3^{(0)}=2\sqrt{\sigma }
\left[ a(n)/3\right] ^{3/4}.$ To calculate the value of $\beta ^{dia}$ for
nucleons we can use the theoretical magnitude of the mean-squared relative
coordinate $\langle {\bf r}^2\rangle $ or experimental data for the size
of a nucleon. The first way is preferable. Taking into account Eq.
(47), the value of $M$ and using the approximate relation $\langle {\bf
\ r}^2\rangle \simeq \langle \sqrt{{\bf r}^2}\rangle ^2,$ Eq. (65)
transforms into
\begin{equation}
\beta ^{dia}=-\frac{e^2+4q^2}{12\sigma ^{3/2}}\left[ \frac{a(n)}3\right]
^{3/4}.  \label{(66)}
\end{equation}

For a proton with the more favorable configuration a), when the quark $
q^{(3)}=d$ and diquark $(q^{(1)}q^{(2)})=(uu),$ parameter $q=(5e)/3$ and Eq.
(5.66) (at $\sigma =0.15$ GeV$^2$) takes the value
\begin{equation}
\beta _p^{dia}=-8\times 10^{-4}~ fm^3.  \label{(67)}
\end{equation}

This quantity is greater than that found in the nonrelativistic quark model
[18,14]. The value of the magnetic polarizability extracted from the
low-energy Compton experiments [19-22] is
\begin{equation}
\bar{\beta }=\beta ^{para}+\beta ^{dia},  \label{(68)}
\end{equation}
where the paramagnetic polarizability is given by [18,14]
\begin{equation}
\beta ^{para}=2\sum_n \frac{\mid \langle n\mid m_z\mid
0\rangle \mid ^2}{{\cal E}_n-{\cal E}_0}.
\label{(69)}
\end{equation}

Here $\mid 0\rangle $, $\mid n\rangle $ are the ground and excited
states of a nucleon, respectively, $m_z$ is the third projection
of the magnetic dipole operator. The main contribution to the
paramagnetic polarizability of a nucleon is due to from $\Delta
(1232)$ excitation [28] and is given by $\beta _\Delta
^{para}=\left( 13\pm 3\right) \times 10^{-4}$ fm$^3$. The
intermediate state $\Delta$ is an isovector excitation and
contributes to the proton and neutron so that
$\beta^p_\Delta=\beta^n_\Delta$. To calculate $\beta^{para}$ in
the present approach one needs to take into account the
interaction of spins of quarks with the magnetic field. For
estimation of the total magnetic polarizability of a proton we use
the contribution $\beta _\Delta ^{para}$ and in accordance with
Eqs. (67), (68) we find
\begin{equation}
\bar{\beta }_p=\left( 5\pm 3\right) \times 10^{-4} fm^3.
\label{(70)}
\end{equation}
 This quantity is close (within two standard deviations) to the experimental value
[22] which was extracted from the measurements of the cross
sections of the Compton scattering on hydrogen:
$\bar{\beta}_p^{\exp }=\left( 2.9\mp 0.7\mp 0.8\right) \times
10^{-4}$ fm$^3$. The value (70) is also in agreement with
$\bar{\beta}_p^{\exp }=\left( 2.1\mp 0.8\mp 0.5\right) \times
10^{-4}$ fm$^3$ [19] and with the quantity found in CHPT [28]. The
value $\bar{\beta}_p$ is small due to the partial cancellation of
the positive paramagnetic polarizability $\bar{\beta }_p^{para}$
with the negative diamagnetic polarizability $\bar{\beta
}_p^{dia}$. The theoretical evaluation, interpretation and
experimental extraction of the total magnetic polarizability
$\bar{\beta }_p$ is difficult because it is small. That is why
this quantity is model dependent in different schemes. The
approach considered allows us to improve the accuracy by using the
perturbation in the spin interaction. The next step is to take
into account such corrections.

For a neutron with the configuration a) where the quark $q^{(3)}=u$ and
diquark $(q^{(1)}q^{(2)})=(dd)$, the parameter $q=-(4e)/3$ and Eq. (65)
gives
\begin{equation}
\beta _n^{dia}=-5.4\times 10^{-4}~ fm^3.
\label{(71)}
\end{equation}

Using the paramagnetic polarizability $\beta _\Delta
^{para}=\left( 13\pm 3\right) \times 10^{-4}$ fm$^3$ [30] for a
neutron we get from (68) the generalized magnetic polarizability
of a neutron
\begin{equation}
\bar{\beta }_n=\left( 7.6\pm 3\right) \times 10^{-4} fm^3
\label{(72)}
\end{equation}
 which is in agreement with the
experimental quantities
\[
\beta _n=\left( 6.5\mp 2.4\mp 3\right) \times 10^{-4}~ fm^3~~~
[23]
\]
\[
\beta _n^{\exp }=\left( 3.2\mp 1.5\mp 2.0\right) \times 10^{-4}~
fm^3~~~ [22]
\]
According to Eqs. (57), (59), (70), (72) the sum of nucleon
polarizabilities in our approach is given by
\[
\bar{\alpha}_p+\bar{\beta }_p=\left(15\pm 3\right) \times 10^{-4}~ fm^3,
\]
\begin{equation}
\bar{\alpha}_n+\bar{\beta }_n=\left(11.8\pm 3\right) \times 10^{-4}~ fm^3.
 \label{(73)}
\end{equation}
Values (73) are close to the reliable results found on the basis
of the sum rule [27]. The interesting feature of the experimental
data is that for the proton and the neutron the approximate
equality $\bar {\alpha}_p+\bar {\beta}_p \simeq \bar {\alpha}_n+
\bar {\beta}_n$, and inequality $\bar {\alpha}_p, \bar
{\alpha}_n\gg \bar {\beta}_p, \bar {\beta}_n$ are valid. The last
inequality means that both a proton and a neutron behave like
electric dipoles. In addition, the large positive paramagnetic
polarizability of nucleons should be cancelled by the diamagnetic
contribution. It was shown in the framework of the sum rule [27]
that in the chiral limit, $m_{\pi}\rightarrow 0$, both
polarizabilities $\bar{\alpha}$, $\bar{\beta}$ diverge as
$O(m_{\pi}^{-1})$, and this was confirmed in CHPT [28]. The same
behavior of pion polarizabilities was found on the basis of
quasiclassical calculations [31] in the framework of the instanton
vacuum theory. Thus, the leading terms of electromagnetic
polarizabilities are determined by CSB. As pion degrees of freedom
play a very important role in the nucleon electric polarizability,
one needs to describe pions in this scheme. In the approach
considered here the accuracy of the values of electromagnetic
polarizabilities can be improved by taking into consideration pion
degrees of freedom.

\section {Conclusion}

The model of baryons as a quark-diquark system is very similar to
the approach for mesons [2]. The effective action for baryons (21)
obtained on the basis of the QCD string theory is applied here for
calculating the mean size and electromagnetic polarizabilities of
a proton and neutron which are in reasonable agreement with the
experimental data. The preferable combination for a diquark is
$(uu)$ for a proton and $(dd)$ for a neutron. In this case
theoretical values for electromagnetic polarizabilities are close
to experimental data.

It should be noted that the quark-diquark ansatz used here generates an electric dipole
moment, and a result, there is a contribution to the energy linear in the electric field
which is like the linear Stark effect.

As a first step, we made some approximations and model
assumptions. So, spin interactions of quarks treated as a
perturbation were neglected here, but we took them into account by
using the paramagnetic polarizability of a nucleon due to the
$\Delta$ contribution. As an approximation, it is justified,
because in Isgur-Karl model of baryons [32] spin-orbit splitting
are much smaller than expected from one-gluon-exchange matrix
elements (spin forces were also discussed in [33]). It was also
shown [34,35] that the contributions from the Coulomb and
spin-spin interactions cancel each other. Nevertheless all
parameters should be taken from the approach and then compared
with empirical data. Therefore one needs to calculate the
paramagnetic polarizability of a nucleon in our approach.

Implying small spin-orbit forces in baryons, we come to $K^{*}$
and quark-diquark system correspondence (see [2]). It is possible
also to consider excited states of baryons at nonzero orbital,
$l$, and principal, $n$, quantum numbers. So, the present approach
can be applied for studying any baryons (see also [34]).

The electromagnetic polarizabilities of nucleons found are close
to the values calculated in the framework of the dispersion sum
rule [27] and CHPT [28].

To improve the accuracy of calculations of electromagnetic
polarizabilities of nucleons, one should take into account pion
degrees of freedom because the pion cloud contributes
substantially to electromagnetic properties.

The nucleon polarizabilities were also evaluated on the basis of
nucleon soliton models in [36-42] .

\end{document}